\documentclass[12pt,a4paper,dvips]{article}
\usepackage{a4p}
\usepackage{cite,mcite}
\usepackage{graphicx,epsfig,ifthen}
\usepackage{physics}
\usepackage{l3_title,Lep}
\usepackage{rotating}
\journalname{Phys. Lett. B}
\date{November 6, 2001}
\preprint{2001-080}
\Lep{2}
\newlength{\capindent}
\newlength{\capwidth}
\newlength{\figwidth}
\newcommand{\icaption}[2][!*!,!]{\hspace*{\capindent}%
  \begin{minipage}{\capwidth}
    \ifthenelse{\equal{#1}{!*!,!}}%
      {\caption{#2}}%
      {\caption[#1]{#2}}
  \end{minipage}}
\def \be {\begin{equation}}
\def \e {\end{equation}}
\def \bea {\begin{eqnarray}}
\def \ea {\end{eqnarray}}

\def \g {\gamma}

\def \EXCALIBUR {{\tt EXCALIBUR}}
\def \EEWWG {{\tt EEWWG}}

\def \KORALW {{\tt KORALW}}
\def \RacoonWW {{\tt RacoonWW}}
\def \yfs {{\tt YFSWW3}}

\def \ipb {~pb$^{-1}$}

\def \wwg {$\Wp\Wm\gamma$ }
\def \nngg {$\nu\bar\nu\gamma\gamma$ }
\def \xs {cross section }

\def \no {\nonumber}

\newcommand{\To}[2]{\stackrel{#1}{\hbox to #2 pt{\rightarrowfill}}}

\begin{document}
\begin{titlepage}
  \title{Study of the \boldmath{$\Wp\Wm\gamma$} Process and Limits on \\
  Anomalous Quartic Gauge Boson Couplings at LEP} 

  \author{The L3 Collaboration}
%
%
\begin{abstract}
  The process $\epem\ra\Wp\Wm\gamma$ is studied using the data
  collected by the L3 detector at LEP. New results, corresponding to an integrated
  luminosity of 427.4\,pb$^{-1}$ at centre-of-mass energies from
  $192 \GeV$ to $207 \GeV$, are presented.

  The $\Wp\Wm\gamma$ cross sections are measured to be in agreement
  with Standard Model expectations.  
No hints of anomalous quartic gauge boson couplings are observed.
Limits at  95\% confidence level are derived using also the  
process $\epem \ra \nu\bar\nu \g\g$.

\end{abstract}

\submitted
\end{titlepage}

\section*{Introduction}
The increase of the LEP centre-of-mass energy well above the W boson
pair-production threshold opens the possibility of studying the triple
boson production process $\epem\ra\Wp\Wm\gamma$. We report
on the  cross section measurement
of this inclusive process where the photon lies inside 
a defined phase space region.

\begin{figure}[p]
\begin{center}
\raisebox{8cm}{\Large \bf 
\begin{tabular}{l}
a) \\ \\ \\ \\ \\ \\ \\ \\ \\ \\ \\ 
b)
\end{tabular}
}~\includegraphics[width=0.5\linewidth]{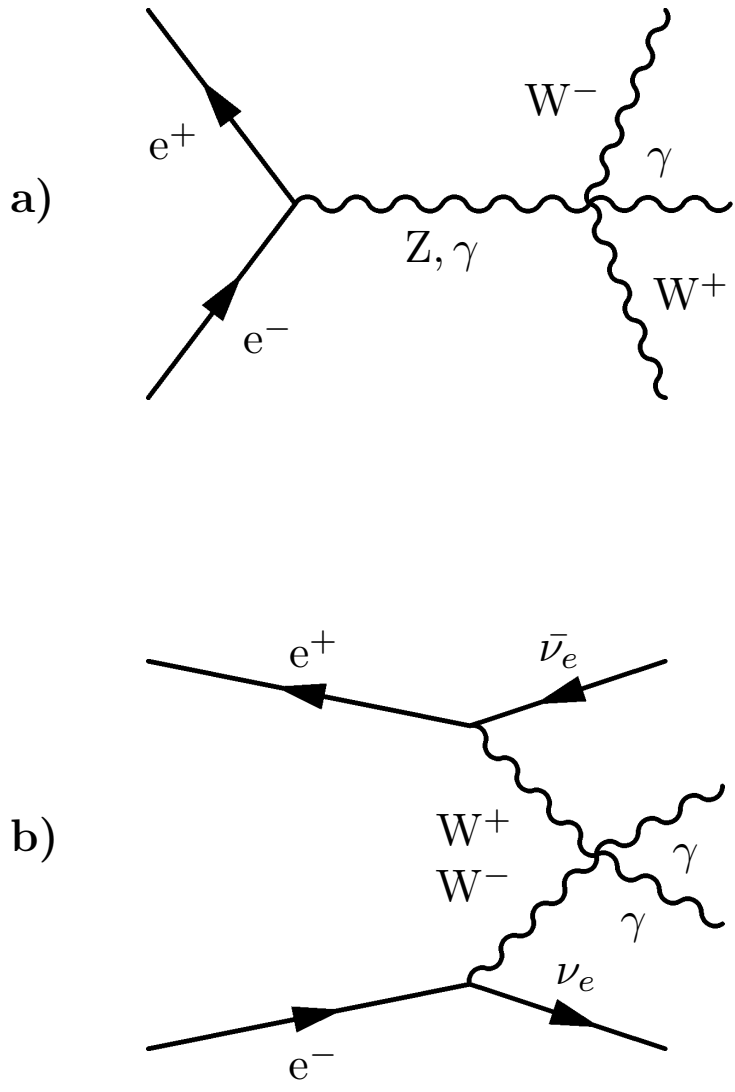}

\caption{Feynman diagrams containing four-boson
  vertices leading to the a) \wwg and b) 
  $\nu\bar\nu \g\g$ final states.}
\label{fey}
\end{center}
\end{figure}

The three boson final state gives access to quartic gauge
boson couplings represented by four-boson interaction vertices 
as shown in Figure~\ref{fey}a.
At the LEP centre-of-mass energies the contribution of four-boson
vertex diagrams, predicted by the  Standard
Model of electroweak interactions~\cite{standard_model,SM-2}, 
are negligible with respect to the other competing
diagrams, mainly initial-state radiation. 
The study of the \wwg process is  thus sensitive to 
anomalous quartic gauge couplings (AQGC)
in both the $\Wp\Wm\Zo\gamma$ and $\Wp\Wm\gamma\gamma$ vertices.
The presence of AQGC
would increase the cross section and modify the photon energy
spectrum of the \wwg process. This search is performed within the theoretical framework of
References~\citen{bud} and~\citen{anja}.

The existence of AQGC would also affect the $\epem
\ra \nu \bar\nu \g\g$ process via the
$\Wp\Wm$ fusion diagram, shown in  Figure~\ref{fey}b, 
containing the $\Wp\Wm\gamma\gamma$
vertex~\cite{stir_phot}.  The
reaction $\epem \ra \nu\bar\nu\g\g$ is dominated by initial-state radiation 
whereas the quartic Standard Model contribution from the $\Wp\Wm$ fusion is
negligible at LEP.  
Also in this case the presence of 
AQGC would enhance the production rate, especially for the
hard tail of the photon energy distribution and for photons produced
at large angles with respect to the beam direction.

The results are based on the high energy data sample collected with
the L3 detector~\cite{l3det}.
Data at the centre-of-mass energy 
of $\sqrt{s}=189 \GeV$, corresponding to an integrated luminosity of
176.8~pb$^{-1}$, were already 
analysed~\cite{articolo} and are used in the 
AQGC analysis.      
In the following, particular emphasis is given to the additional
luminosity of 427.4~pb$^{-1}$ recorded at
centre-of-mass energies ranging from $\sqrt{s}= 192 \GeV$ up to  $207 \GeV$. 

The results derived on AQGC from the \wwg and \nngg
channels are eventually combined. 

Studies of triple gauge bosons production and AQGC were recently
reported for both the
$\Wp\Wm\gamma$~\cite{opal-qgc} and 
$\Zo\gamma\gamma$~\cite{zgg} final states.

%
%
\section*{Monte Carlo Simulation}
The dominant contributions to the 
$\Wp\Wm\gamma$ final state come from the radiative graphs with photons
emitted  by the incoming particles (ISR), by
the decay products of the W bosons (FSR) or by the W's themselves (WSR).

In this study, the signal is defined as the phase space region of the 
$\epem\ra\Wp\Wm\gamma$ process where the photon fulfills the following
criteria:

\begin{itemize}
\item $E_\gamma >$ 5 GeV, where $E_\gamma$ is the energy of the
  photon,
\item $20^\circ < \theta_\gamma < 160^\circ$, where $\theta_\gamma$ is the angle
  between the photon and the beam axis,
\item $\alpha_\gamma >$ 20$^\circ$, where $\alpha_\gamma$ is the angle
  between the direction of the photon and that of the closest charged fermion.
\end{itemize}

These requirements, used to enhance the effect of possible AQGC, 
largely contribute to avoid infrared and collinear
singularities  in the calculation of the signal cross section.

In order to study efficiencies, background contaminations 
and AQGC effects, several Monte Carlo programs are used.  

The \KORALW \cite{KORALW} generator, which does not include 
either the quartic coupling diagrams or the WSR, performs initial state 
multi-photon radiation  in the full photon phase space. FSR from
charged leptons in the event up to double bremsstrahlung is included
using the {\tt PHOTOS}~\cite{photos} package.  
The {\tt JETSET}~\cite{js} Monte Carlo program, which includes photons
in the parton shower, is used to model the fragmentation and
hadronization process.
The \KORALW\ program is used
in the analysis for the determination of efficiencies. 
{\tt PYTHIA}~\cite{pyt}  is used to simulate the
background processes: ${\rm e^+e^-}\ra \Zo/\gamma\ra {\rm q
  \bar q}(\gamma)$, ${\rm e^+e^-}\ra \Zo\Zo\ra 4\mathrm{f}\kern 0.05em(\gamma)$ and
${\rm e^+e^-}\ra \Zo\mathrm{ee}\ra \mathrm{f}\kern 0.05em \mathrm{f}\kern 0.05em\mathrm{ee}(\gamma)$.

The \EEWWG\ \cite{anja} program is used to simulate the effect of AQGC.
It includes ${\cal O}(\alpha)$ calculations for visible photons
but is lacking the simulation of photons collinear to the beam pipe
and of FSR.
The net effect of collinear photons, included by 
implementing the \EXCALIBUR \cite{exca}
collinear radiator function,  is to move the effective 
centre-of-mass energy towards lower values, reducing the expected
signal cross section by about 18\%.

Other Monte Carlo programs which include WSR or full ${\cal
  O}(\alpha)$ corrections, such as \yfs\ \cite{YFSWW3} and
  \RacoonWW\ \cite{RACOONWW}, are used to cross check the calculations.

For the  simulation of the $\epem \ra \nu\bar\nu\g\g$ process in the
framework of the Standard Model the {\tt KORALZ}~\cite{koralz_new}  Monte Carlo
generator is used.
{\tt NUNUGPV} \cite{NUNUGPV_new} is
also used to cross check the results, and found to be in agreement
with {\tt KORALZ}. 
The effects of AQGC 
are simulated using the {\tt EENUNUGGANO} program~\cite{stir_phot}.
The missing higher order corrections due to
ISR in  {\tt EENUNUGGANO} are also  estimated
 by implementing the {\tt
EXCALIBUR} collinear radiator function.

The response of the L3 detector is
modelled with the GEANT~\cite{geant} detector simulation program
which includes the effects of energy loss, multiple scattering and
showering in the detector materials and in the beam pipe.  
Time dependent detector inefficiencies are taken into account in the
simulation.

\section*{\boldmath{$\Wp\Wm\gamma$} Event Selection and Cross Section}
The selection of $\Wp\Wm\gamma$ events follows two steps:
first semileptonic $\mathrm{W^+W^- \rightarrow qqe}\nu$ or
qq$\mu\nu$, and fully hadronic $\mathrm{W^+W^-\rightarrow qqqq}$
events are selected~\cite{l3-pre98}, then a
search for isolated photons is performed.  

The photon identification in $\Wp\Wm$ events is optimized for each
four-fermion final state. Photons are identified from  energy
clusters in the electromagnetic calorimeter not associated with any
track in the central detector and with low activity in  the nearby
region of the hadron calorimeter.
The profile of the shower must be consistent with that of an
electromagnetic particle.  
Experimental cuts on photon energy and angles are applied, reflecting the phase
space definition of the signal.

\begin{figure}[p]
\begin{center}
  \includegraphics[width=0.96\linewidth]{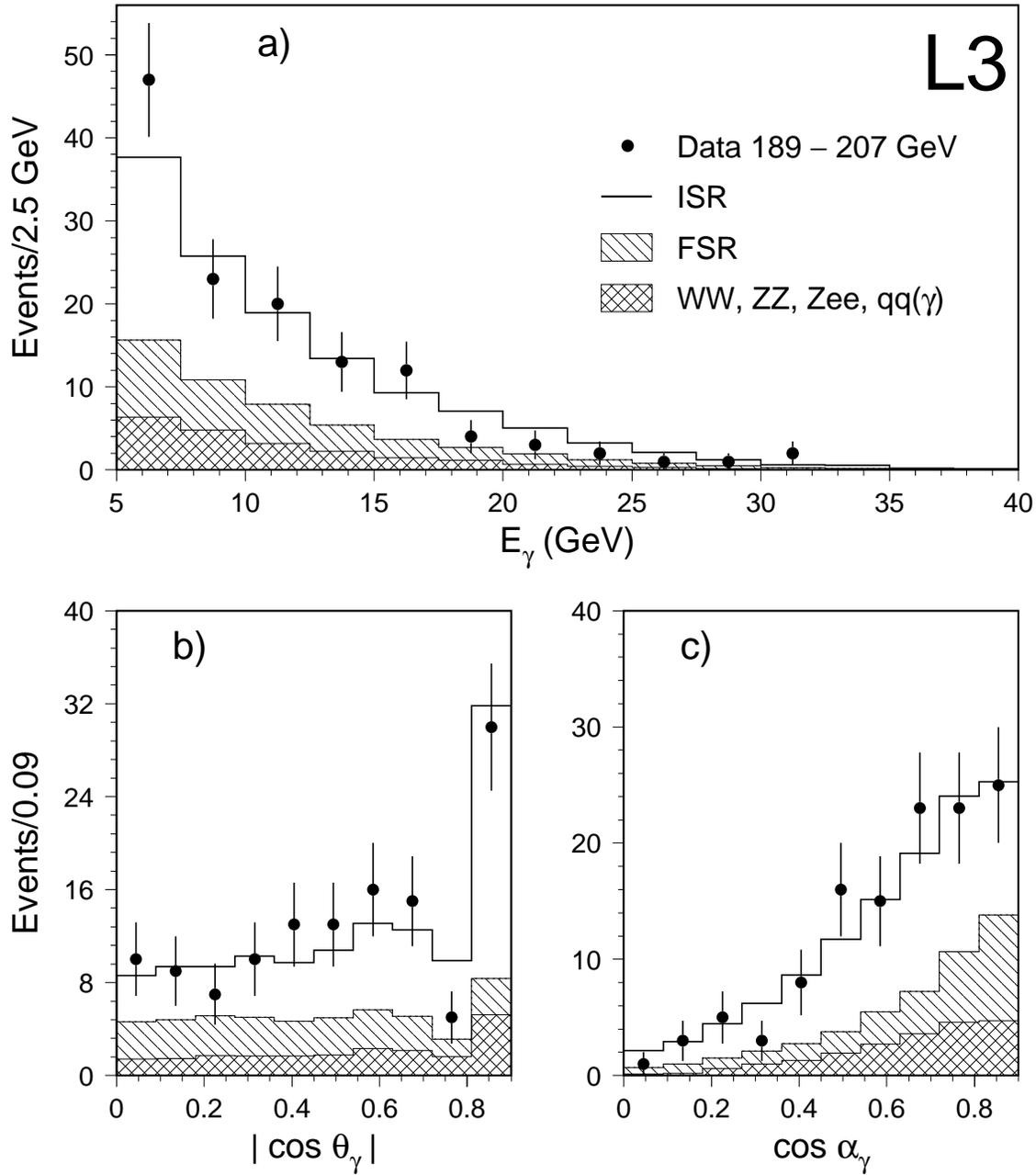}
  \caption[]{Distributions at the centre-of-mass
    energies $\rts=189 - 207 \GeV$ of a) the photon energy, b) the
    cosine of the angle of the photon to  
    the beam axis and c) the cosine of the angle of the photon to the
    closest charged lepton or jet.}
\label{spec189}
\end{center}
\end{figure}

\begin{figure}[p]
\begin{center}
  \includegraphics[width=0.95\linewidth]{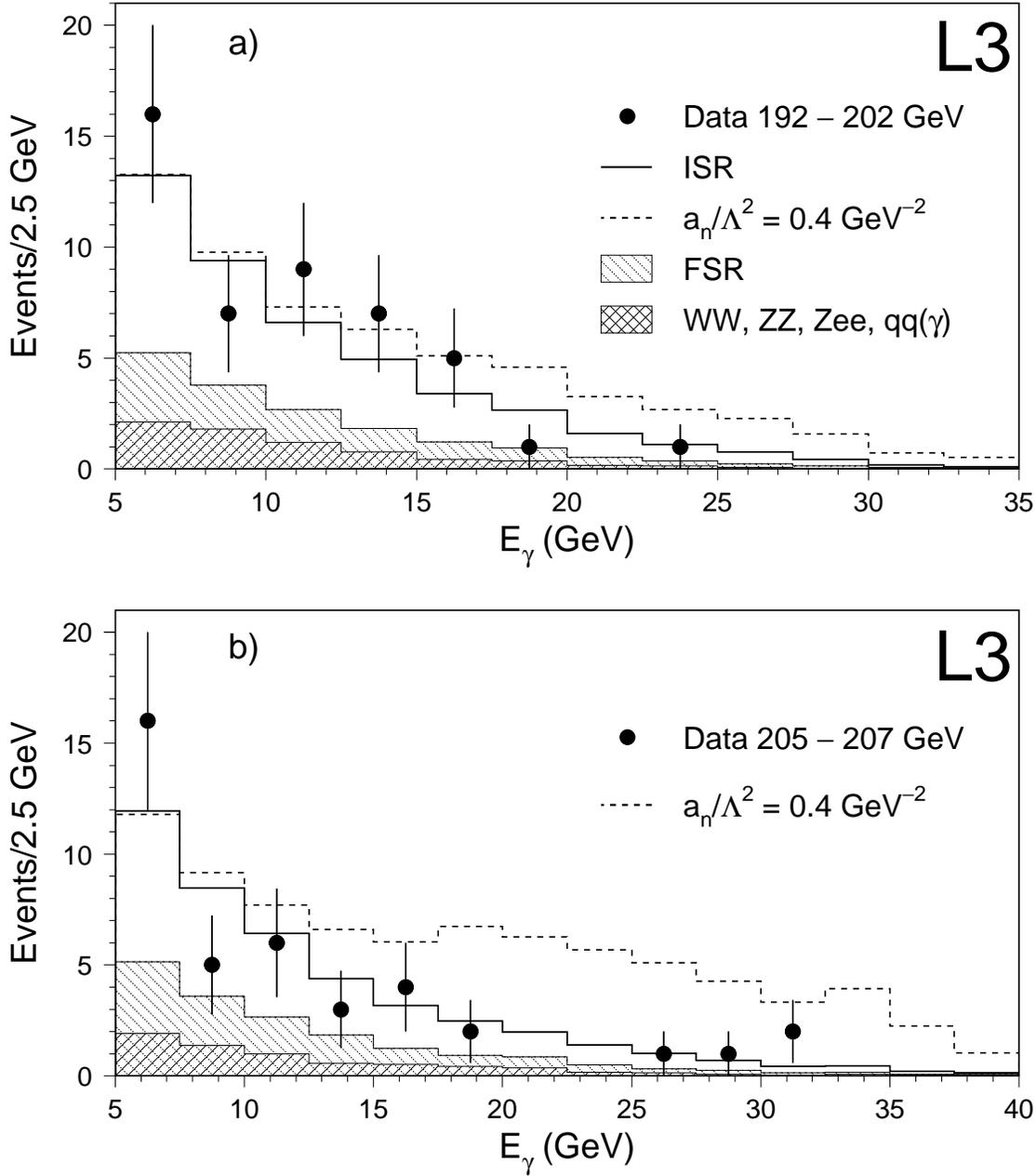}
\caption[]{Distributions of the photon energy
  for the semileptonic qqe$\nu$, qq$\mu\nu$ and fully hadronic
  $\Wp\Wm\gamma$ decay modes corresponding to the data collected at
  a) $\sqrt{s}=192 - 202 \GeV$ and b) $\sqrt{s}=205 - 207 \GeV$.  
  The cross-hatched area is the background
  component from WW, ZZ, Zee, and ${\rm q \bar q(\gamma)}$ events. The  FSR
  distribution includes the contribution of photons radiated off
  the charged fermions and photons originating from isolated meson
  decays.  Distributions corresponding to non-zero values of the
  anomalous coupling $a_n/\Lambda^2$ are shown as dashed lines. }
\label{fig:sele}
\end{center}
\end{figure}

Figure~\ref{spec189}
shows the distributions of $E_\gamma$, $\theta_\gamma$, and
$\alpha_\gamma$ for the full data set, including the data at
$\sqrt{s}=189 \GeV$.
Here, $\alpha_\gamma$ is defined as the angle of the photon with
respect to the closest identified lepton or hadronic jet. 
Good agreement between data and 
Standard Model expectations is observed.
Figure~\ref{fig:sele} shows the distributions of $E_\gamma$ for the
data collected at $\sqrt{s}=192 - 202 \GeV$ and $\sqrt{s}=205
-207 \GeV$, respectively. 
  
Table~\ref{tab:wwg} summarizes the selection yield. 
In total 86 $\Wp\Wm\gamma$ candidate events are selected at
$\sqrt{s}=192-207 \GeV$. The Standard Model
expectation, inside the specified phase space region, is of 
$87.8\pm0.8$ events.  

The quantity $\varepsilon_{\rm WW}$, representing the selection 
efficiencies for the $\mathrm{W^+W^-}\rightarrow$~qql$\nu$ and qqqq decay
modes, ranges from 70\% to 87\%. 
The quantity $\varepsilon_{\gamma}$ is the photon identification
efficiency  inside the selected phase space region. This efficiency
takes into account small effects of events migrating from outside
the signal region into the selected sample due to the finite detector
resolution. Its value ranges from 52\% to 80\%, the lowest
efficiencies being obtained in the fully hadronic sample where the
high multiplicity makes the
photon identification more difficult.
The overall selection efficiency   
$\varepsilon_{\rm WW}\times\varepsilon_{\g}$ is around
45\% for all final states.

The $\Wp\Wm\gamma$ cross sections are evaluated channel by channel and
then combined according to the Standard Model W boson branching
fractions.
The data samples at $\rts = 192 -196 \GeV$, $\rts = 200 - 202 \GeV$
and $\rts = 205 - 207  \GeV $ are respectively 
merged. They correspond to the luminosity averaged centre-of-mass
energies and to the integrated
luminosities listed in Table~\ref{tab:wwg}.

\begin{figure}[p]
\begin{center}

  \includegraphics[width=0.96\linewidth]{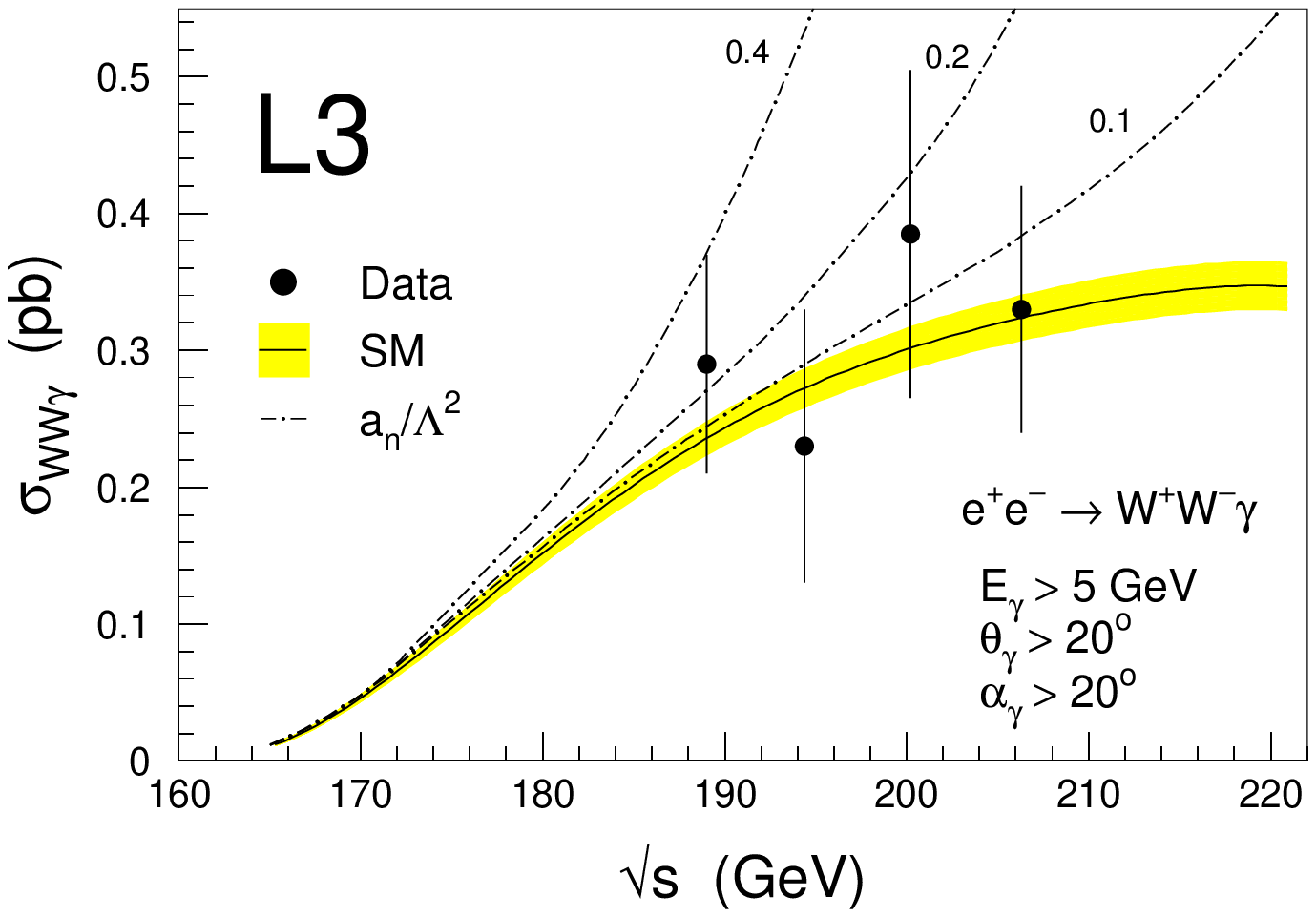}
  
  \vspace*{-1cm}
  \caption[]{
    Measured \xs for the process $\epem\ra\Wp\Wm\gamma$ compared to
    the Standard Model \xs as a function of the centre-of-mass energy, as
    predicted by the \EEWWG\ Monte Carlo within 
    phase-space requirements.  The shaded band corresponds to a theoretical
    uncertainty of $\pm$5\%.  The three dash-dotted lines correspond to the
    \xs for the indicated values of the anomalous coupling
    $a_n/\Lambda^2$ (in GeV$^{-2}$ units).\\
}

\label{fig:xsec}
\end{center}
\end{figure}

The results,
including the published value at $\rts=189\GeV$~\cite{articolo}, are:
\begin{eqnarray*}
\sigma_{{\rm WW}\gamma}(188.6 \GeV) & =  & 0.29 \pm 0.08 \pm 0.02
~~{\rm pb}~~~~{\rm (\sigma_\mathrm{SM} = 0.233 \pm 0.012 \;\;pb)} \\ 
\sigma_{{\rm WW}\gamma}(194.4 \GeV) & =  & 0.23 \pm 0.10 \pm 0.02
~~{\rm pb}~~~~{\rm (\sigma_\mathrm{SM} = 0.268 \pm 0.013\;\;pb)} \\ 
\sigma_{{\rm WW}\gamma}(200.2 \GeV) & =  & 0.39 \pm 0.12 \pm 0.02
~~{\rm pb}~~~~{\rm (\sigma_\mathrm{SM} = 0.305 \pm 0.015\;\;pb)} \\ 
\sigma_{{\rm WW}\gamma}(206.3 \GeV) & =  & 0.33 \pm 0.09 \pm 0.02
~~{\rm pb}~~~~{\rm (\sigma_\mathrm{SM} = 0.323 \pm 0.016\;\;pb)}, 
\end{eqnarray*}
where the first uncertainty is statistical and the second systematic.
The measurements are in good agreement with 
the Standard Model expectations, $\sigma_\mathrm{SM}$, 
calculated using \EEWWG\ and
reported with a theoretical uncertainty of 5\%~\cite{priv}.
Figure~\ref{fig:xsec} shows these results together with the
predicted total $\Wp\Wm\gamma$ cross sections 
as a function of the centre-of-mass energy.

The ratio between the measured
cross section $\sigma_{\rm meas}$ and the theoretical expectations
is derived at each centre-of-mass energy. 
These values are then combined as:
$$ R=\frac{\sigma_{\rm meas}}{\sigma_{\rm SM}} = 1.09 \pm 0.17 \pm
0.09\; , $$
where the first uncertainty is statistical and the second systematic.

The systematic uncertainties arising in the inclusive W-pair event
selections~\cite{l3-pre98} are propagated to the final measurement and
correspond to an uncertainty of 0.008 pb
for all the energy points.  Additional
systematic uncertainties due to the electromagnetic
calorimeter resolution and  energy scale
are found to be negligible.
The total systematic uncertainty is dominated by the {\tt JETSET}
modelling of photons from meson decays ($\pi^0, \eta$).  
Its effect has been directly studied on data~\cite{articolo} comparing
the photon rate in ${\rm e^+e^-}\ra \Zo \ra {\rm q
  \bar q}(\gamma)$ events with Monte Carlo simulations. A correction factor
of $ 1.2 \pm 0.1$ is applied to the rate of photons in the Monte Carlo simulation 
and its uncertainty is propagated.
This uncertainty, fully correlated among the data taking periods,
amounts to 6~\% of the measured cross section.

%
\section*{Determination of Anomalous Quartic Gauge Couplings}

\subsection*{The \boldmath$\epem\ra\Wp\Wm\gamma$ Process}

In the framework of References~\citen{bud} and~\citen{anja}, the Standard Model Lagrangian of
electroweak interactions is extended to include dimension-6 operators 
proportional to the three AQGC: 
$a_0/\Lambda^2$, $a_c/\Lambda^2$ and
$a_n/\Lambda^2$, where $\Lambda^2$ represents the energy
scale for new physics.

The two parameters $a_0/\Lambda^2$ and
$a_c/\Lambda^2$, which are separately C and P conserving, generate
anomalous $\Wp\Wm\gamma\gamma$ and $\Zo\Zo\gamma\gamma$ vertices. The
term $a_n/\Lambda^2$, which is CP violating, gives rise to an
anomalous contribution to the $\Wp\Wm\Zo\gamma$ vertex.  Indirect limits on
$a_0/\Lambda^2$ and $a_c/\Lambda^2$ were derived~\cite{indi}, 
but only the study of $\Wp\Wm\gamma$
events allows for a direct measurement of the anomalous coupling
$a_n/\Lambda^2$.
These couplings would manifest themselves by modifying the energy
spectrum of the photons and the total cross section as shown in
Figures~\ref{fig:sele} and~\ref{fig:xsec}, respectively.
The effect increases with  increasing centre-of-mass energy.
These predictions are obtained by reweighting the \KORALW\ Monte Carlo
events by the ratio of the differential distributions as calculated by the 
\EEWWG\ and \KORALW\ programs~\cite{articolo}.

The derivation of AQGC is performed by fitting both the 
shape and the normalization of the photon energy spectrum in the 
range from 5 \GeV ~to 35 \GeV. Each of the AQGC
is varied in turn fixing the other two to zero. 

The combination of all data, including the results at $\sqrt{s}=189\GeV$~\cite{articolo}, 
gives:
\bea
a_0/\Lambda^2  &=& \phantom{-} 0.000\pm0.010 \GeV^{-2}\no \\
a_c/\Lambda^2  &=& - 0.013\pm0.023 \GeV^{-2}\no \\
a_n/\Lambda^2  &=& -0.002\pm0.076 \GeV^{-2}\no 
\ea
where systematic uncertainties are included.

At the 95\% confidence level, the AQGC are constrained to:
\begin{eqnarray*}  
-0.017  \GeV^{-2} & < & a_0/\Lambda^2 ~ < ~ 0.017 \GeV^{-2} \\ 
-0.052  \GeV^{-2} & < & a_c/\Lambda^2 ~ < ~ 0.026 \GeV^{-2} \\ 
-0.14\phantom{0} \GeV^{-2} & < & a_n/\Lambda^2 ~ < ~ 0.13\phantom{0} \GeV^{-2}.
\end{eqnarray*}
All these results are in agreement with the Standard Model
expectation. The sign of the $a_0$ and $a_c$ AQGC, obtained with the
\EEWWG\ reweighting, is reversed according  to the discussions in
References~\citen{RACWWG} and \citen{WRAP}.

\subsection*{The \boldmath$\epem \ra \nu\bar\nu\g\g$ Process }

The sensitivity of the $\epem \ra \nu\bar\nu\g\g$ process to the
$a_0/\Lambda^2$ and $ a_c/\Lambda^2$ AQGC, 
through the diagrams shown in Figure~\ref{fey}b, is also 
exploited. Events with an acoplanar multi-photon signature are selected~\cite{papl3gg}.
In this letter we report on results from data at $\rts =192 - 207 \GeV$.

\begin{figure}[p]
\begin{center}
\includegraphics[width=0.85\linewidth]{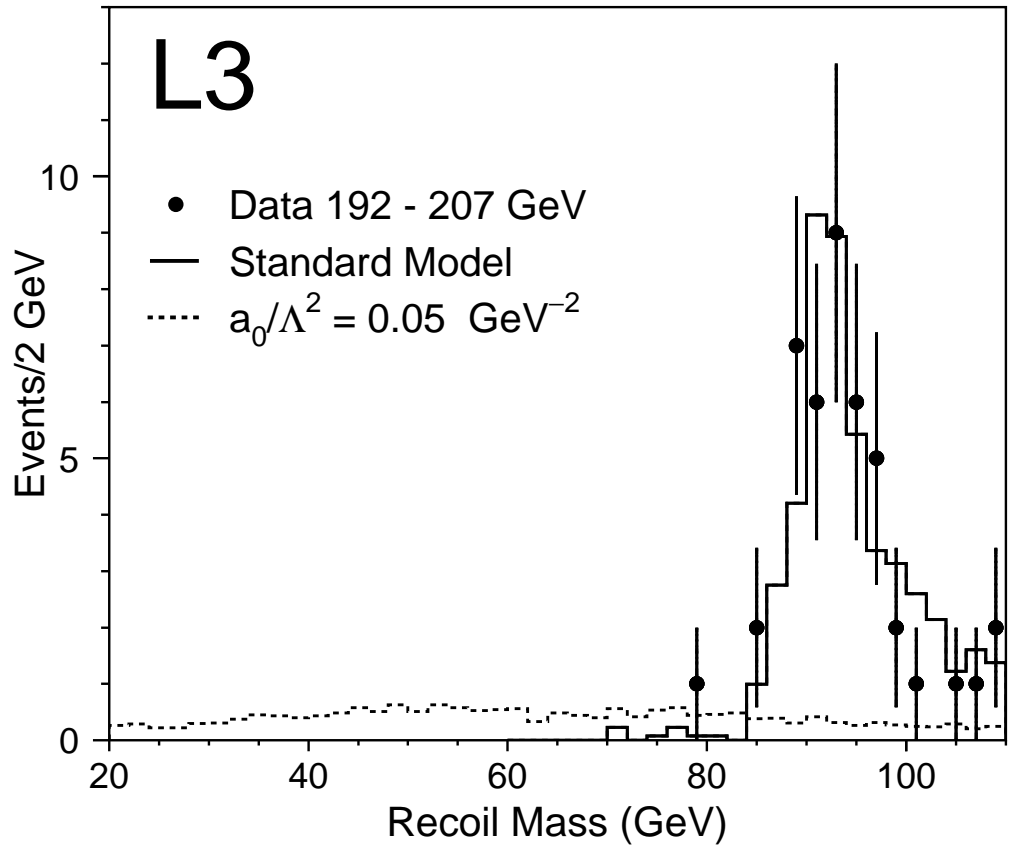} 
\caption{ Recoil mass spectrum of the acoplanar photon pair events
  selected at $\sqrt{s}= 192-207\GeV$. }
\label{fig:recmass}
\end{center}
 \end{figure}

Figure~\ref{fig:recmass} shows the
two-photon recoil mass, $M_\mathrm{rec}$, distribution, with the predicted AQGC signal
for a non-zero anomalous coupling $a_0/\Lambda^2$. The number of
selected events in the Z peak region, defined as $75 \GeV < M_\mathrm{rec} < 110
\GeV$,  is 43 in agreement with the Standard
Model expectation of $47.6 \pm 0.7$.  

The AQGC signal prediction is reliable only for recoil masses lower than the mass of
the Z boson as the interference with the Standard Model
processes is not included in the calculation. 
Requiring $M_\mathrm{rec}<75\GeV$, no event is retained by the selection 
in agreement with the Standard Model expectation of $0.35\pm0.05$ events. 

A reweighting technique based on
the full matrix elements as calculated by the {\tt EENUNUGGANO} Monte
Carlo,
is used to derive the AQGC values.
Including the results at $\rts=183\GeV$ and
$\rts=189\GeV$~\cite{articolo}, 
the 95\% confidence level upper limits are:
\begin{eqnarray*} 
-0.031 \GeV^{-2} & < & a_0/\Lambda^2 ~ < ~ 0.031 \GeV^{-2}\\ 
-0.090 \GeV^{-2}  & < & a_c/\Lambda^2 ~ < ~ 0.090 \GeV^{-2}\,,
\end{eqnarray*}
where only one parameter is varied at a time.

The dominant systematic uncertainty comes from the theoretical
uncertainty of 5\%~\cite{priv} in the calculation of anomalous cross
sections.

\subsection*{Combined Results}

The results obtained from the $\Wp\Wm\gamma$ and $\nu\bar\nu
\g\g$ processes are combined.
No evidence of AQGC is found  and 95\% confidence level limits are
obtained separately on each coupling as: 
\begin{eqnarray*} 
-0.015  \GeV^{-2}          & < ~ a_0/\Lambda^2 & < ~ 0.015 \GeV^{-2} \\ 
-0.048  \GeV^{-2}          & < ~ a_c/\Lambda^2 & < ~ 0.026 \GeV^{-2} \\ 
-0.14\phantom{0} \GeV^{-2} & < ~ a_n/\Lambda^2 & < ~ 0.13\phantom{0} \GeV^{-2}.
\end{eqnarray*}

\section*{Appendix}
The results on
the $\epem \ra \Wp\Wm\gamma$ cross sections 
are also expressed for a different phase space region
defined by:
\begin{itemize}
\item $E_\gamma >$ 5 GeV,
\item $|\cos(\theta_\gamma)| <$ 0.95,
\item $\cos(\alpha_\gamma) <$ 0.90,
\item $M_{ff'} = M_{\rm W} \pm 2 \Gamma_{\rm W}$, where $M_{ff'}$ are
  the two fermion-pair invariant masses. 
\end{itemize}
The results read:
\begin{eqnarray*}
\sigma_{{\rm WW}\gamma}(188.6 \GeV) & =  & 0.20 \pm 0.09 \pm 0.01
~~{\rm pb}~~~~{\rm (\sigma_\mathrm{SM} = 0.190 \pm 0.010\;\; pb)} \\ 
\sigma_{{\rm WW}\gamma}(194.4 \GeV) & =  & 0.17 \pm 0.10 \pm 0.01
~~{\rm pb}~~~~{\rm (\sigma_\mathrm{SM} = 0.219 \pm 0.011\;\; pb)} \\ 
\sigma_{{\rm WW}\gamma}(200.2 \GeV) & =  & 0.43 \pm 0.13 \pm 0.02
~~{\rm pb}~~~~{\rm (\sigma_\mathrm{SM} = 0.242 \pm 0.012\;\; pb)} \\ 
\sigma_{{\rm WW}\gamma}(206.3 \GeV) & =  & 0.13 \pm 0.08 \pm 0.01
~~{\rm pb}~~~~{\rm (\sigma_\mathrm{SM} = 0.259 \pm 0.013\;\; pb),} 
\end{eqnarray*}
where the first uncertainty is statistical, the second systematic and the values
 in parentheses indicate the Standard Model predictions. 

%
%
\newpage
\section*{Author List}
\typeout{   }     
\typeout{Using author list for paper 248 }
\typeout{Using author list for paper 248 ONLY ! }
\typeout{Using author list for paper 248 ONLY ! }
\typeout{Using author list for paper 248 ONLY ! }
\typeout{$Modified: Jul 31 2001 by smele $}
\typeout{!!!!  This should only be used with document option a4p!!!!}
\typeout{   }
%
%
%
%
%
%

\newcount\tutecount  \tutecount=0
\def\tutenum#1{\global\advance\tutecount by 1 \xdef#1{\the\tutecount}}
\def\tute#1{$^{#1}$}
\tutenum\aachen            
\tutenum\nikhef            
\tutenum\mich              
\tutenum\lapp              
\tutenum\basel             
\tutenum\lsu               
\tutenum\beijing           
\tutenum\berlin            
\tutenum\bologna           
\tutenum\tata              
\tutenum\ne                
\tutenum\bucharest         
\tutenum\budapest          
\tutenum\mit               
\tutenum\panjab            
\tutenum\debrecen          
\tutenum\florence          
\tutenum\cern              
\tutenum\wl                
\tutenum\geneva            
\tutenum\hefei             
\tutenum\lausanne          
\tutenum\lyon              
\tutenum\madrid            
\tutenum\florida           
\tutenum\milan             
\tutenum\moscow            
\tutenum\naples            
\tutenum\cyprus            
\tutenum\nymegen           
\tutenum\caltech           
\tutenum\perugia           
\tutenum\peters            
\tutenum\cmu               
\tutenum\potenza           
\tutenum\prince            
\tutenum\riverside         
\tutenum\rome              
\tutenum\salerno           
\tutenum\ucsd              
\tutenum\sofia             
\tutenum\korea             
\tutenum\utrecht           
\tutenum\purdue            
\tutenum\psinst            
\tutenum\zeuthen           
\tutenum\eth               
\tutenum\hamburg           
\tutenum\taiwan            
\tutenum\tsinghua          

{
\parskip=0pt
\noindent
{\bf The L3 Collaboration:}
\ifx\selectfont\undefined
 \baselineskip=10.8pt
 \baselineskip\baselinestretch\baselineskip
 \normalbaselineskip\baselineskip
 \ixpt
\else
 \fontsize{9}{10.8pt}\selectfont
\fi
\medskip
\tolerance=10000
\hbadness=5000
\raggedright
\hsize=162truemm\hoffset=0mm
\def\r{\rlap,}
\noindent

P.Achard\r\tute\geneva\ 
O.Adriani\r\tute{\florence}\ 
M.Aguilar-Benitez\r\tute\madrid\ 
J.Alcaraz\r\tute{\madrid,\cern}\ 
G.Alemanni\r\tute\lausanne\
J.Allaby\r\tute\cern\
A.Aloisio\r\tute\naples\ 
M.G.Alviggi\r\tute\naples\
H.Anderhub\r\tute\eth\ 
V.P.Andreev\r\tute{\lsu,\peters}\
% ...
\newpage
\begin{list}{A}{\itemsep=0pt plus 0pt minus 0pt\parsep=0pt plus 0pt minus 0pt
                \topsep=0pt plus 0pt minus 0pt}
\item[\aachen]
 I. Physikalisches Institut, RWTH, D-52056 Aachen, FRG$^{\S}$\\
 III. Physikalisches Institut, RWTH, D-52056 Aachen, FRG$^{\S}$
\item[\nikhef] National Institute for High Energy Physics, NIKHEF, 
     and University of Amsterdam, NL-1009 DB Amsterdam, The Netherlands
\item[\mich] University of Michigan, Ann Arbor, MI 48109, USA
\item[\lapp] Laboratoire d'Annecy-le-Vieux de Physique des Particules, 
     LAPP,IN2P3-CNRS, BP 110, F-74941 Annecy-le-Vieux CEDEX, France
\item[\basel] Institute of Physics, University of Basel, CH-4056 Basel,
     Switzerland
\item[\lsu] Louisiana State University, Baton Rouge, LA 70803, USA
\item[\beijing] Institute of High Energy Physics, IHEP, 
  100039 Beijing, China$^{\triangle}$ 
\item[\berlin] Humboldt University, D-10099 Berlin, FRG$^{\S}$
\item[\bologna] University of Bologna and INFN-Sezione di Bologna, 
     I-40126 Bologna, Italy
\item[\tata] Tata Institute of Fundamental Research, Mumbai (Bombay) 400 005, India
\item[\ne] Northeastern University, Boston, MA 02115, USA
\item[\bucharest] Institute of Atomic Physics and University of Bucharest,
     R-76900 Bucharest, Romania
\item[\budapest] Central Research Institute for Physics of the 
     Hungarian Academy of Sciences, H-1525 Budapest 114, Hungary$^{\ddag}$
\item[\mit] Massachusetts Institute of Technology, Cambridge, MA 02139, USA
\item[\panjab] Panjab University, Chandigarh 160 014, India.
\item[\debrecen] KLTE-ATOMKI, H-4010 Debrecen, Hungary$^\P$
\item[\florence] INFN Sezione di Firenze and University of Florence, 
     I-50125 Florence, Italy
\item[\cern] European Laboratory for Particle Physics, CERN, 
     CH-1211 Geneva 23, Switzerland
\item[\wl] World Laboratory, FBLJA  Project, CH-1211 Geneva 23, Switzerland
\item[\geneva] University of Geneva, CH-1211 Geneva 4, Switzerland
\item[\hefei] Chinese University of Science and Technology, USTC,
      Hefei, Anhui 230 029, China$^{\triangle}$
\item[\lausanne] University of Lausanne, CH-1015 Lausanne, Switzerland
\item[\lyon] Institut de Physique Nucl\'eaire de Lyon, 
     IN2P3-CNRS,Universit\'e Claude Bernard, 
     F-69622 Villeurbanne, France
\item[\madrid] Centro de Investigaciones Energ{\'e}ticas, 
     Medioambientales y Tecnol\'ogicas, CIEMAT, E-28040 Madrid,
     Spain${\flat}$ 
\item[\florida] Florida Institute of Technology, Melbourne, FL 32901, USA
\item[\milan] INFN-Sezione di Milano, I-20133 Milan, Italy
\item[\moscow] Institute of Theoretical and Experimental Physics, ITEP, 
     Moscow, Russia
\item[\naples] INFN-Sezione di Napoli and University of Naples, 
     I-80125 Naples, Italy
\item[\cyprus] Department of Physics, University of Cyprus,
     Nicosia, Cyprus
\item[\nymegen] University of Nijmegen and NIKHEF, 
     NL-6525 ED Nijmegen, The Netherlands
\item[\caltech] California Institute of Technology, Pasadena, CA 91125, USA
\item[\perugia] INFN-Sezione di Perugia and Universit\`a Degli 
     Studi di Perugia, I-06100 Perugia, Italy   
\item[\peters] Nuclear Physics Institute, St. Petersburg, Russia
\item[\cmu] Carnegie Mellon University, Pittsburgh, PA 15213, USA
\item[\potenza] INFN-Sezione di Napoli and University of Potenza, 
     I-85100 Potenza, Italy
\item[\prince] Princeton University, Princeton, NJ 08544, USA
\item[\riverside] University of Californa, Riverside, CA 92521, USA
\item[\rome] INFN-Sezione di Roma and University of Rome, ``La Sapienza",
     I-00185 Rome, Italy
\item[\salerno] University and INFN, Salerno, I-84100 Salerno, Italy
\item[\ucsd] University of California, San Diego, CA 92093, USA
\item[\sofia] Bulgarian Academy of Sciences, Central Lab.~of 
     Mechatronics and Instrumentation, BU-1113 Sofia, Bulgaria
\item[\korea]  The Center for High Energy Physics, 
     Kyungpook National University, 702-701 Taegu, Republic of Korea
\item[\utrecht] Utrecht University and NIKHEF, NL-3584 CB Utrecht, 
     The Netherlands
\item[\purdue] Purdue University, West Lafayette, IN 47907, USA
\item[\psinst] Paul Scherrer Institut, PSI, CH-5232 Villigen, Switzerland
\item[\zeuthen] DESY, D-15738 Zeuthen, 
     FRG
\item[\eth] Eidgen\"ossische Technische Hochschule, ETH Z\"urich,
     CH-8093 Z\"urich, Switzerland
\item[\hamburg] University of Hamburg, D-22761 Hamburg, FRG
\item[\taiwan] National Central University, Chung-Li, Taiwan, China
\item[\tsinghua] Department of Physics, National Tsing Hua University,
      Taiwan, China
\item[\S]  Supported by the German Bundesministerium 
        f\"ur Bildung, Wissenschaft, Forschung und Technologie
\item[\ddag] Supported by the Hungarian OTKA fund under contract
numbers T019181, F023259 and T024011.
\item[\P] Also supported by the Hungarian OTKA fund under contract
  number T026178.
\item[$\flat$] Supported also by the Comisi\'on Interministerial de Ciencia y 
        Tecnolog{\'\i}a.
\item[$\sharp$] Also supported by CONICET and Universidad Nacional de La Plata,
        CC 67, 1900 La Plata, Argentina.
\item[$\triangle$] Supported by the National Natural Science
  Foundation of China.
\end{list}
}
\vfill


\newpage
%

\bibliographystyle{l3stylem}

\begin{mcbibliography}{10}

\bibitem{standard_model}
S.~L. Glashow, \NP {\bf 22} (1961) 579; S. Weinberg, \PRL {\bf 19} (1967) 1264;
  A. Salam, in {\em Elementary Particle Theory}, ed. N. Svartholm, Stockholm,
  Alm\-quist and Wiksell (1968), 367\relax
\relax
\bibitem{SM-2}
M.~Veltman, \NP {\bf B7} (1968) 637; G.M.~'t~Hooft, \NP {\bf B35} (1971) 167;
  G.M.~'t~Hooft and M.~Veltman, \NP {\bf B44} (1972) 189; \NP {\bf B50} (1972)
  318\relax
\relax
\bibitem{bud}
G. B\'elanger and F. Boudjema,
\newblock  Nucl. Phys. {\bf B288}  (1992) 201\relax
\relax
\bibitem{anja}
J.W. Stirling and A. Werthenbach, \EPJ {\bf C14} (2000) 103\relax
\relax
\bibitem{stir_phot}
J.W.~Stirling and A.~Werthenbach,
\newblock  Phys. Lett. {\bf B466}  (1999) 369\relax
\relax
\bibitem{l3det}
L3 Collab., B.~Adeva $\etal$, Nucl. Instr. Meth. {\bf A289} (1990) 35;
  J.~A.~Bakken $\etal$, Nucl. Instr. Meth. {\bf A275} (1989) 81; O.~Adriani
  $\etal$, Nucl. Instr. Meth. {\bf A302} (1991) 53; B.~Adeva $\etal$, Nucl.
  Instr. Meth. {\bf A323} (1992) 109; K.~Deiters $\etal$, Nucl. Instr. Meth.
  {\bf A323} (1992) 162; M.~Chemarin $\etal$, Nucl. Instr. Meth. {\bf A349}
  (1994) 345; M.~Acciarri $\etal$, Nucl. Instr. Meth. {\bf A351} (1994) 300;
  G.~Basti $\etal$, Nucl. Instr. Meth. {\bf A374} (1996) 293; A.~Adam $\etal$,
  Nucl. Instr. Meth. {\bf A383} (1996) 342\relax
\relax
\bibitem{articolo}
L3 Collab., M. Acciarri \etal,
\newblock  Phys. Lett. {\bf B490}  (2000) 187\relax
\relax
\bibitem{opal-qgc}
Opal Collab., G. Abbiendi {\em et al.},
\newblock  Phys. Lett. {\bf B471}  (1999) 293\relax
\relax
\bibitem{zgg}
L3 \coll, M. Acciarri \etal, \PL {\bf B478} (2000) 34; Phys.\ Lett. {\bf B505}
  (2001) 47\relax
\relax
\bibitem{KORALW}
KORALW version 1.33 is used,\\ S. Jadach \etal, Comp. Phys. Comm. {\bf 94}
  (1996) 216; S. Jadach \etal, Phys. Lett. {\bf B372} (1996) 289\relax
\relax
\bibitem{photos}
E. Barberio, B. van Eijk and Z. Was, { Comp. Phys. Comm.} {\bf 79} (1994)
  291\relax
\relax
\bibitem{js}
JETSET version 7.409 is used,\\ T. Sj{\"o}strand and H. U. Bengtsson, {Comp.
  Phys. Comm.} {\bf 46} (1987) 43\relax
\relax
\bibitem{pyt}
PYTHIA version 5.722 is used,\\ T. Sj{\"o}strand, {Comp. Phys. Comm.} {\bf 82}
  (1994) 74\relax
\relax
\bibitem{exca}
F.A. Berends, R. Pittau and R. Kleiss, {Comp. Phys. Comm.} {\bf 85} (1995)
  437\relax
\relax
\bibitem{YFSWW3}
YFSWW3 version 1.14 is used, \\ S.~Jadach \etal, \PR {\bf D54} (1996) 5434;
  Phys. Lett. {\bf B417} (1998) 326; \PR {\bf D61} (2000) 113010; preprint
  hep-ph/0007012 (2000)\relax
\relax
\bibitem{RACOONWW}
A.~Denner \etal, \PL {\bf B475} (2000) 127; Nucl.\ Phys. {\bf B587} (2000)
  67.\relax
\relax
\bibitem{koralz_new}
KORALZ version 4.03 is used,\\ S. Jadach {\em et al.}, \CPC {\bf 79} (1994)
  503\relax
\relax
\bibitem{NUNUGPV_new}
G. Montagna {\em et al.},
\newblock  Nucl. Phys. {\bf B541}  (1999) 31\relax
\relax
\bibitem{geant}
GEANT Version 3.15 is used,\\ R. Brun \etal, ``GEANT 3'', preprint CERN
  DD/EE/84-1 (1984), revised 1987.\\ The GHEISHA program (H. Fesefeldt, RWTH
  Aachen Report PITHA 85/02 (1985)) is used to simulate hadronic
  interactions\relax
\relax
\bibitem{l3-pre98}
L3 Collab., M. Acciarri \etal,
\newblock  Phys. Lett. {\bf B496}  (2000) 19\relax
\relax
\bibitem{priv}
J.W. Stirling and A. Werthenbach, private communication\relax
\relax
\bibitem{indi}
O.J.P. Eboli and M.C. Gonzales--Garcia, Phys. Lett. {\bf B411} (1994) 381\relax
\relax
\bibitem{RACWWG}
A. Denner \etal, Eur.\ Phys.\ J. {\bf C20} (2001) 201\relax
\relax
\bibitem{WRAP}
G. Montagna \etal, Phys.\ Lett. {\bf B515} (2001) 197\relax
\relax
\bibitem{papl3gg}
L3 \coll, M. Acciarri \etal, \PL {\bf B444} (1998) 503; \PL {\bf B470} (1999)
  268\relax
\relax
\end{mcbibliography}

\newpage

\vspace*{5cm}
\begin{table}[htbp]
\begin{center}
  \renewcommand{\arraystretch}{1.5}
\begin{tabular}{|c||c|c|c|c|c|c|}
\hline
$\langle\sqrt{s}\rangle$ & ~~Channel~~ &$N_{\it data}$& $\varepsilon_{\rm WW}$ 
& $\varepsilon_{\gamma}$ &
$N^{exp}_{\it TOT}$ & $N^{exp}_{\it Bkgr}$ \\
\hline\hline
194.4 \GeV &qqe$\nu\gamma$   & $\phantom{0}4$ &
$0.748\pm0.007$ & $0.613\pm0.023$  & $\phantom{0}3.76 \pm0.14 $ & $\phantom{0} 1.41\pm0.09 $ \\
(113.4\ipb) &qq$\mu\nu\gamma$ & $\phantom{0}6$ & 
$0.731\pm0.007$ & $0.728\pm0.026$  & $\phantom{0}4.52 \pm0.15 $ & $\phantom{0}1.68 \pm0.09 $ \\
           &qqqq$\gamma$     & $\phantom{0}9$ & 
$0.865\pm0.004$ & $0.537\pm0.013$  & $12.89\pm0.34 $ & $\phantom{0}5.22 \pm0.28 $ \\
\hline\hline                                                               
 200.2 \GeV  &qqe$\nu\gamma$ & $\phantom{0}5$ & 
$0.726\pm0.007$ & $0.631\pm0.023$  & $\phantom{0}4.28 \pm0.14 $ & $\phantom{0}1.47 \pm0.08 $ \\
(119.8\ipb)  &qq$\mu\nu\gamma$ & $\phantom{0}4$ & 
$0.712\pm0.007$ & $0.744\pm0.025$  & $\phantom{0}5.50 \pm0.16 $ & $\phantom{0}2.07 \pm0.10 $ \\
             &qqqq$\gamma$     & $18          $ & 
$0.836\pm0.004$ & $0.521\pm0.012$  & $13.71\pm0.29 $ & $\phantom{0}5.31 \pm0.28 $ \\
\hline\hline                                                               
206.3 \GeV   &qqe$\nu\gamma$   & $\phantom{0}7$ & 
$0.700\pm0.005$ & $0.626\pm0.024$  & $\phantom{0}6.58 \pm0.22 $ & $\phantom{0}2.21 \pm0.13 $ \\
(194.2\ipb)  &qq$\mu\nu\gamma$ & $\phantom{0}4$ & 
$0.714\pm0.005$ & $0.787\pm0.026$  & $\phantom{0}9.54 \pm0.26 $ & $\phantom{0}3.48 \pm0.16 $ \\
             &qqqq$\gamma$     & $29          $ & 
$0.823\pm0.005$ & $0.540\pm0.011$  & $26.97\pm0.50 $ & $12.11\pm0.39 $ \\
\hline
\end{tabular}
\caption[]{Number of observed events, $N_{\it data}$,
  W$^+$W$^-$ and W$^+$W$^-\gamma$ selection
  efficiencies, $\varepsilon_{\rm WW}$ and $\varepsilon_{\gamma}$,  
  expected total number of events, $N^{exp}_{\it TOT}$, and
  background estimates, $N^{exp}_{\it Bkgr}$, for the various decay channels according to the
  Standard Model prediction.   The background estimates include
  FSR and misidentified photons.
  All uncertainties come from Monte Carlo
  statistics. The average centre-of-mass energies, $\langle\sqrt{s}\rangle$, and the integrated
  luminosity of the three subsamples  are also listed.
}
\label{tab:wwg}
\end{center}
\end{table}

\newpage

\end{document}